\documentclass[12pt]{article}
\usepackage[english]{babel}
\usepackage{cite}
\usepackage{amsmath,amssymb}
\usepackage{epsfig}

\begin{document}
\thispagestyle{empty}

\vspace*{2.0cm}

\begin{center}
\boldmath
 {\large \bf
HYPERDIRE \\
HYPERgeometric functions DIfferential REduction: \\
Mathematica-based packages for the differential reduction of generalized
hypergeometric functions: \\
Lauricella function $F_C$ of three variables
 }
\unboldmath
\end{center}
 \vspace*{0.8cm}

\begin{center}
{\sc Vladimir~V.~Bytev,$^{a,b,}$\footnote{E-mail: bvv@jinr.ru}
Bernd~A.~Kniehl$^{a,}$\footnote{E-mail: kniehl@desy.de}} \\
 \vspace*{1.0cm}
{\normalsize $^{a}$ II. Institut f\"ur Theoretische Physik, Universit\"at Hamburg,}\\
{\normalsize Luruper Chaussee 149, 22761 Hamburg, Germany} \\
\bigskip
{\normalsize $^{b}$ Joint Institute for Nuclear Research,} \\
{\normalsize $141980$ Dubna (Moscow Region), Russia}
\end{center}

\begin{abstract}
We present a further extension of the HYPERDIRE project, which is devoted to
the creation of a set of Mathematica-based program packages for manipulations
with  Horn-type hypergeometric functions on the basis of differential
equations.
Specifically, we present the implementation of the differential reduction for
the Lauricella function $F_C$ of three variables.
\end{abstract}

\newpage

{\bf\large PROGRAM SUMMARY}
\vspace{4mm}
\begin{sloppypar}
\noindent   
   {\em Program title\/}: HYPERDIRE \\[2mm]
   {\em Version\/}: 3.0.0 \\[2mm]
   {\em Release\/}: 1.0.0 \\[2mm]
   {\em Catalogue identifyer\/}: \\[2mm]
   {\em Program summary URL\/}: \\[2mm]
   {\em Program obtainable from\/}:
        {\tt https://sites.google.com/site/loopcalculations/home} \\[2mm]
   {\em Licensing provisions\/}: GNU General Public Licence \\[2mm]
   {\em No.\ of lines in distributed program, including test data etc.\/}:
   \\[2mm]
   {\em No.\ of bytes in distributed program, including test data etc.\/}:
   \\[2mm]
   {\em Distribution format\/}: {\tt tar.gz} \\[2mm]
   {\em Programming language\/}: Mathematica. \\[2mm]
   {\em Computer\/}: All computers running Mathematica. \\[2mm]
   {\em Operating systems\/}: Operating systems running Mathematica. \\[2mm]
   {\em Classification\/}: \\[2mm]
   {\em Does the new version supersede the previous version?\/}: No, it
        significantly extends the previous version. \\[2mm]
   {\em Nature of the problem\/}: Reduction of hypergeometric function $F_C$ of
        three variables to a set of basis functions. \\[2mm]
   {\em Solution method\/}: Differential reduction. \\[2mm]
   {\em Restriction on the complexity of the problem}: None. \\[2mm]
   {\em Reasons for new version\/}: The extension package allows the user to
        handle the Lauricella function $F_C$ of three variable. \\[2mm]
   {\em Summary of revisions\/}: The previous version goes unchanged. \\[2mm]
   {\em Running time}: Depends on the complexity of the problem.
\end{sloppypar}
%
\newpage



\section{Introduction}

Multiloop and/or multileg Feynman diagrams as well as phase space integrals in
covariant gauge within dimensional regularization \cite{Bollini:1972ui} can
be written in terms of generalized hypergeometric functions.
The creation of the HYPERDIRE program packages
\cite{hyperdire1,hyperdire2,hyperdire3,hyperdire4} is motivated by the
importance of Horn-type hypergeometric functions for the analytical evaluations
of Feynman diagrams, especially at the one-loop level \cite{one-loop}.
Possible applications of the differential-reduction algorithm to Feynman
diagrams beyond the one-loop level were discussed in Ref.~\cite{our}.

 
A Feynman diagram may be written in the form of a Mellin-Barnes integral
\cite{MB}, which depends on external kinematic invariants, the dimension $n$ of
space-time, and the powers of the propagators.
Upon application of Cauchy's theorem, the Feynman integral can be converted
into a linear combination of multiple series:
\begin{equation}
\Phi(n,\vec{x}\,) \sim
\sum_{k_1,\cdots, k_{r+m}=0}^\infty
\prod_{a,b}
\frac{\Gamma(\sum_{i=1}^m {A}_{ai}k_i+{B}_a)}{\Gamma(\sum_{j=1}^r
{C}_{bj}k_j+{D}_{b})}
x_1^{k_1} \cdots x_{r+m}^{k_{r+m}}\;,
\label{Horn}
\end{equation}
where $x_i$ are some rational functions of Mandelstam variables and 
$A_{ai}, B_a, C_{bj}, D_b$ are linear functions of the space-time dimension and
the propagator powers.
The representation of Eq.~(\ref{Horn}) corresponds to a Horn-type
hypergeometric series \cite{Gelfand} if the hidden index of the summation is
considered as an independent variable.

In general, the multiple series
\begin{equation}
H(\vec{z}) = \sum_{\vec{m}=0}^\infty C(\vec{m}) \vec{z}^{\,\,\vec{m}}\;,
\label{horn1}
\end{equation}
where $\vec{m}=(m_1,\ldots,m_r)$ and
$\vec{z}^{\,\,\vec{m}}=z_1^{m_1}\cdots z_r^{m_r}$, are called
{\it Horn-type hypergeometric} if, for each $i=1, \ldots, r$, the ratio
$C(\vec{m}+\vec{e}_j)/C(\vec{m})$, where 
$\vec{e}_j = (0,\cdots,0,1,0,\cdots,0)$ is the $j^\mathrm{th}$ unit vector, is a
rational function of the summation indices, i.e.\
\begin{equation}
\frac{C(\vec{m}+\vec{e}_j)}{C(\vec{m})}  =  \frac{P_j(\vec{m})}{Q_j(\vec{m})}
\;,
\label{horn2}
\end{equation}
where $P_j(\vec{m})$ and $Q_j(\vec{m})$ are polynomials \cite{Gelfand,bateman}.
In explicit form, the coefficients $C(\vec{m})$ can then be written as
\begin{equation}
C(\vec{m})
= \prod_{i=1}^r \lambda_i^{m_i}R(\vec{m})
\frac{
\prod_{j=1}^N \Gamma(\vec{\mu_j}\cdot \vec{m}+\gamma_j)
}
{
\prod_{k=1}^M \Gamma(\vec{\nu_k}\cdot \vec{m}+\delta_k)
}\;,
\label{ore}
\end{equation}
where $N, M \geq 0$, $\lambda_i,\gamma_j,\delta_k$ are arbitrary complex
numbers, $\vec{\mu_j}$, $\vec{\nu_k}$ are arbitrary integer-valued vectors,
and $R$ is an arbitrary rational function.

From the condition (\ref{horn2}) on the coefficients $C(\vec{m})$ of the
Horn-type hypergeometric function $H(\vec{z})$, we can derive the following
proper system of partial differential equations (PDEs):
\begin{equation}
\left[
Q_j(
\vec{\theta})
\frac{1}{z_j}
-
 P_j(
 \vec{\theta}
)
\right]
H(\vec{z})=0 \;,
\label{diff}
\end{equation}
where $j=1, \ldots, r$, $\vec{\theta}=(\theta_1,\ldots,\theta_r)$, and
$\theta_k$ is the differential operator 
\begin{equation}
\theta_k=z_k\frac{\partial}{\partial z_k}\;.
\end{equation}

In previous publications \cite{hyperdire1,hyperdire2,hyperdire3,hyperdire4}, we
presented the Mathematica-based \cite{math} package HYPERDIRE for the
differential reduction of Horn-type hypergeometric functions.
In Ref.~\cite{hyperdire1}, we implemented the reduction of the Horn-type
hypergeometric functions ${}_{p+1}F_{p}$ of one variable to restricted sets of
basis functions and predicted the numbers of such functions.
We demonstrated that the differential-reduction algorithm can be used for the
reduction of  Feynman diagrams without resorting to the integration-by-parts
technique.
We established and implemented the criterion of reducibility of the Horn-type
hypergeometric functions ${}_{p+1}F_{p}$ to simpler functions for special
values of parameters.
Subsequently, we developed the HYPERDIRE project further to cover the full set
of Horn-type hypergeometric functions of two variables \cite{hyperdire3},
including the Appell functions $F_1$, $F_2$, $F_3$, and $F_4$
\cite{hyperdire2}, and also certain Horn-type hypergeometric functions of three
variables, namely $F_D$ and $F_S$ \cite{hyperdire4}.
In this paper, we discuss the case of the Horn-type hypergeometric function
$F_C$ of three variable, which appears, e.g., in the calculation of two-loop
bubble-type Feynman diagram  with different masses.
With the implementation of $F_C$, $F_D$, and $F_S$, we start to study the
applicability of the differential-reduction method to the set of
Lauricella--Saran hypergeometric functions of three variables.

\section{Differential reduction of Horn-type hypergeometric functions}

Let us now consider the Horn-type hypergeometric function
$H(\vec{z})=H(\vec{\gamma};\vec{\sigma};\vec{z})$, which explicitly depends on
a set of {\it contiguous} variables, $\vec{z}=(z_1, \ldots, z_k)$, and two
sets of {\it discrete} variables, $\vec{\gamma}=(\gamma_1,\ldots, \gamma_i)$
and $\vec{\sigma}=(\sigma_1,\ldots, \sigma_j)$, which are called {\it upper}
and {\it lower} parameters, respectively.

In Refs.~\cite{SST,theorem}, it was shown that there exist unique linear
differential operators which can generate identities called contiguous or
ladder relations between the hypergeometric function
$H(\vec{\gamma};\vec{\sigma};\vec{z})$ and its counterparts with one of the
upper (lower) parameters shifted by unity, namely
\begin{eqnarray}
H(\vec{\gamma}+\vec{e_c};\vec{\sigma};\vec{x})
& = &
\frac{1}{\gamma_c}
\left ( \sum_{a=1}^r \mu_{ca} \theta_a+\gamma_c \right)
H(\vec{\gamma};\vec{\sigma};\vec{x})
=U^+_{\gamma_c}H(\vec{\gamma};\vec{\sigma};\vec{x})
\;,
\label{do1}\\
H(\vec{\gamma};\vec{\sigma}-\vec{e}_c;\vec{x})
& = &
\frac{1}{\sigma_c - 1}
\left(
\sum_{b=1}^r
\nu_{cb} \theta_b + \sigma_c - 1
\right)
H(\vec{\gamma};\vec{\sigma};\vec{x})
=U^-_{\sigma_c}H(\vec{\gamma};\vec{\sigma};\vec{x})
\;.
\label{do2}
\end{eqnarray}
The direct operators $U^+_{\gamma_c}$ and $U^-_{\sigma_c}$ are called
{\it step-up} and {\it step-down} operators for the upper and lower indices,
respectively.
It is possible to construct the inverse differential operators $U^-_{\gamma_c}$
and $U^+_{\sigma_c}$ satisfying
\begin{eqnarray}
\bigl[ U^-_{\gamma_c}U^+_{\gamma_c} \bigr] H(\vec{\gamma};\vec{\sigma};\vec{x})
=H (\vec{\gamma};\vec{\sigma};\vec{x})\;,
\nonumber\\
\bigl[U^+_{\sigma_c}U^-_{\sigma_c}\bigr]H(\vec{\gamma};\vec{\sigma};\vec{x}) =
H (\vec{\gamma};\vec{\sigma};\vec{x})\;.
\label{invdef}
\end{eqnarray}
Once these operators are constructed, we can combine them to shift the
parameters of the Horn-type hypergeometric function by any integer, i.e.\ to
obtain contiguous relations of the form
\begin{eqnarray}
\bigl[ U^- U^+ \bigr] H(\vec{\gamma};\vec{\sigma};\vec{x})
=H (\vec{\gamma}+\vec{k};\vec{\sigma}+\vec{l};\vec{x})\;.
\end{eqnarray}
The process of applying $U_{\gamma_c}^\pm$ and $U_{\sigma_c}^\pm$ to a Horn-type
hypergeometric function to shift its parameters by integers is called
{\it differential reduction}.
In this way, the Horn-type structure provides an opportunity to reduce
hypergeometric functions to a set of basis functions with parameters differing
from the original values by integer shifts,
\begin{equation}
H(\vec{\gamma}+\vec{k};\vec{\sigma}+\vec{l};\vec{x})
=
\prod_{i,j,m,n} U^{+}_{\gamma_i}U^{-}_{\gamma_j}U^{+}_{\sigma_m}U^{-}_{\sigma_n}
H(\vec{\gamma};\vec{\sigma};\vec{x}) \;.
\label{reductionX}
\end{equation}

The development of systematic techniques for the solution of contiguous
relations has a long history.
It was started by Gauss, who described the reduction of the hypergeometric
function ${}_2F_1$ in 1823 \cite{Gauss}.
Numerous papers have since then been published on this problem
\cite{contiguous}.
An algorithmic solution was found by Takayama in Ref.~\cite{theorem},
and this method was later extended in a series of publications \cite{takayama}
(see also Refs.~\cite{Ore}).

Previously, it was pointed out \cite{our} that the differential-reduction
algorithm in Eq.~(\ref{reductionX}) can be applied to the reduction of Feynman
diagrams to some subsets of basis hypergeometric functions with well-known
analytical properties and that the system of differential equations in
Eq.~(\ref{diff}) can also be used for the construction of so-called
$\varepsilon$ expansions of hypergeometric functions about rational values of
their parameters via direct solutions of the systems of differential
equations.

\cite{hyperdire4}.  

\boldmath
\section{Lauricella function $F_C$}
\unboldmath
The Lauricella function $F_C$ of three variables \cite{Lauricella} is defined
as a Taylor expansion about the point $\vec{z}=\vec{0}$ as follows:
\begin{eqnarray}
F_C^{(3)}(a,b;c_1,c_2,c_3; z_1,z_2,z_3)
= \sum_{m_1,m_2,m_3=0}^\infty
\frac{(a)_{m_1+m_2+m_3} (b)_{m_1+m_2+m_3}}
     {(c_1)_{m_1} (c_2)_{m_2}(c_3)_{m_3}}\, 
\frac{z_1^{m_1}  z_2^{m_2}  z_3^{m_3}}{m_1! m_2! m_3!}
\;,
\label{FC:series}
\end{eqnarray}
where $(a)_m=(a+m-1)!/(a-1)!$ is the Pochhammer symbol.
The corresponding PDEs of Eq.~(\ref{diff}) read:
\begin{eqnarray}
\label{thForm}
\frac{1}{z_i}\theta_i \left(c_i-1 + \theta_i \right) F_C(\vec{z}) =
\left(a + \theta_1+ \theta_2+\theta_3 \right)
\left(b + \theta_1+ \theta_2+\theta_3 \right) F_C(\vec{z})
\;,
\qquad i = 1,2,3 \;,
\end{eqnarray}
where we have used the short-hand notation
$F_C(\vec{z})=F_C^{(3)}(a,b;c_1,c_2,c_3; z_1,z_2,z_3)$.
The canonical form of Eq.~(\ref{thForm}) reads:
\begin{eqnarray}
\theta_1^2 F_C(\vec{z}) &=&
\frac{1}{D_0}\left\{ -a b z_1
+[(a+b)z_1+(z_2+z_3-1)(c_1-1)]\theta_1
\vphantom{\sum_{i\in\{2,3\}}}
\right.
\nonumber\\
&&{}-\left.
z_1\sum_{i\in\{2,3\}}(1+a+b-c_i)\theta_i
+z_1\sum_{i\ne j}^3 \theta_i\theta_j
\right\} F_C(\vec{z})
\;,
\nonumber
\\
\theta_2^2 F_C(\vec{z}) &=&
\frac{1}{D_0}\left\{ -a b z_2
+[(a+b)z_2+(z_1+z_3-1)(c_2-1)]\theta_2
\vphantom{\sum_{i\in\{2,3\}}}
\right.
\nonumber\\
&&{}-\left.
z_2\sum_{i\in\{1,3\}}(1+a+b-c_i)\theta_i
+z_2\sum_{i\ne j}^3 \theta_i\theta_j
\right\} F_C(\vec{z})
\;,
\nonumber
\\
\theta_3^2 F_C(\vec{z}) &=&
\frac{1}{D_0}\left\{ -a b z_3
+[(a+b)z_3+(z_1+z_3-1)(c_3-1)]\theta_3
\vphantom{\sum_{i\in\{2,3\}}}
\right.
\nonumber\\
&&{}-
\left.
z_3\sum_{i\in\{1,2\}}(1+a+b-c_i)\theta_i
+z_3\sum_{i\ne j}^3 \theta_i\theta_j
\right\} F_C(\vec{z})
\;,
\end{eqnarray}
where $D_0=1-z_1-z_2-z_3$, which can be written in compact form as
\begin{equation}
L_i F_C(\vec{z})=
\theta_i^2 F_C(\vec{z}) =
\left(
\sum_{i\ne j=1}^3P_{ij} \theta_i \theta_j 
+ \sum_{m=1}^3  R_{im} \theta_m
+ S_i
\right) F_C(\vec{z})
\;,
\qquad
i = 1, 2, 3\;. 
\label{canonical}
\end{equation}
Here, we can define the conditions of complete integrability,
\begin{equation}
\theta_i \left[ \theta_j L_k  \right] F_C(\vec{z})
=
\theta_j \left[ \theta_i L_k  \right] F_C(\vec{z})
\;,
\qquad
i,j,k = 1,2,3\;.
\label{integrability}
\end{equation}
Eq.~(\ref{integrability}) does not provide new independent conditions between
the differential operators $\theta_i$.
Thus Eq.~(\ref{canonical}) can be reduced to the following Pfaff system of
eight independent differential equations:
\begin{equation}
d \vec{f} = R \vec{f} \;,
\label{Pfaff}
\end{equation}
where
$
\vec{f} = \left(
F_C(\vec{z}),
\theta_1 F_C(\vec{z}),
\theta_2 F_C(\vec{z}),
\theta_3 F_C(\vec{z}),
\theta_1 \theta_2 F_C(\vec{z}),
\theta_1 \theta_3 F_C(\vec{z}),
\theta_2 \theta_3 F_C(\vec{z}),
\theta_1 \theta_2 \theta_3 F_C(\vec{z})
\right)
$.

\boldmath
\subsection{Differential reduction of $F_C$}
\unboldmath
In the case of the Lauricella function $F_C(\vec{z})$, the direct
differential operators for the upper parameters in Eq.~(\ref{do1}) read:
\begin{eqnarray}
F_C(a+1,b;c_1,c_2,c_3; \vec{z})
&=&U^+_aF_C(\vec{z})=\frac{1}{a}
\left(
a+ \theta_1+\theta_2+\theta_3
\right) F_C(\vec{z})
\;,
\nonumber \\
F_C(a,b+1;c_1,c_2,c_3; \vec{z})
&=&U^+_bF_C(\vec{z})
=\frac{1}{b}
\left(
b+ \theta_1+\theta_2+\theta_3
\right) F_C(\vec{z})
\;,
\label{dirAB}
\end{eqnarray}
and those for the lower parameters in Eq.~(\ref{do2}) read:
\begin{eqnarray}
F_C(a,b;c_1-1,c_2,c_3; \vec{z})
&=&U^-_{c_1}F_C(\vec{z})=\frac{1}{c_1-1}
\left(
c_1 -1 + \theta_1
\right)
F_C(\vec{z}) \;,
\nonumber \\
F_C(a,b;c_1,c_2-1,c_3; \vec{z})
&=&U^-_{c_2}F_C(\vec{z})=
\frac{1}{c_2-1}
\left(
c_2 -1 + \theta_2
\right)
F_C(\vec{z}) \;,
\nonumber \\
F_C(a,b;c_1,c_2,c_3-1; \vec{z})
&=&U^-_{c_3}F_C(\vec{z})=
\frac{1}{c_3-1}
\left(
c_3 -1 + \theta_3
\right)
F_C(\vec{z}) \;.
\label{direct:FS}
\end{eqnarray}
As explained above, we can determine the corresponding inverse differential
operators, $U^-_a$ and $U^-_b$, through eight independent solutions of
Eq.~(\ref{Pfaff}),
\begin{eqnarray}
F_C(a-1,b;c_1,c_2,c_3; \vec{z})
&=&U^-_aF_C(\vec{z})\;,
\nonumber\\
F_C(a,b-1;c_1,c_2,c_3; \vec{z})
&=&U^-_bF_C(\vec{z})\;,
\end{eqnarray}
where 
\begin{equation}
U^-_x=
A_x
+
B_x \theta_1
+
C_x \theta_2
+
D_x \theta_3
+
E_x \theta_{1}\theta_2
+
F_x \theta_{1}\theta_3
+
G_x \theta_{2}\theta_3
+
H_x \theta_{1}\theta_2\theta_3\;,
\label{invOp}
\end{equation}
with $x=a,b$.
Similar solutions can be obtained for the inverse differential operators
$U^+_{c_i}$  ($i=1,2,3$) with eight independent functions in the form of 
Eq.~(\ref{invOp}).

By using the definitions of the inverse operators in Eqs.~(\ref{invdef}) and
(\ref{invOp}), we can explicitly obtain the following equation for the
coefficients
$\vec{A}_x=(A_x,B_x,C_x,D_x,E_x,F_x,G_x,H_x)$:
\begin{eqnarray}
f_0(\vec{A}_x)+\sum_{i=1}^3 f_i(\vec{A}_x)\theta_{i}
+\sum_{i,j=1}^3 f_{ij}(\vec{A}_x)\theta_i\theta_j
+\sum_{i \ne j=1}^3 f_{ijj}(\vec{A}_x)\theta_i\theta_j^2
+ f_{123}(\vec{A}_x)\theta_1\theta_2\theta_3
\nonumber\\
{}+\sum_{i=1}^3 f_{123i}(\vec{A}_x)\theta_1\theta_2\theta_3\theta_i=1\;,
\label{invOpEq}
\end{eqnarray}
where the coefficients $f_{\dots}(\vec{A}_x)$ are linear maps of $\vec{A}_x$ and
rational functions of the discrete and continuous variables of $F_C(\vec{z})$.

Multiplying the PDEs for $F_C(\vec{z})$ in Eq.~(\ref{canonical}) by different
powers of $\theta_i$, we can eliminate higher powers of $\theta_i$ in
Eq.~(\ref{invOpEq}) and write it using a minimal set of eight independent
terms,
\begin{eqnarray}
F_0(\vec{A}_x)+\sum_{i=1}^3 F_i(\vec{A}_x)\theta_i
+\sum_{i \ne j=1}^3 F_{ij}(\vec{A}_x)\theta_i\theta_j
+ F_{123}(\vec{A}_x)\theta_1\theta_2\theta_3=1\;.
\label{invOpEqFin}
\end{eqnarray}
Setting in turn $F_0(\vec{A}_x)=1$, $F_i(\vec{A}_x)=0$, $F_{ij}(\vec{A}_x)=0$,
and $F_{123}(\vec{A}_x)=0$, we obtain eight equations for the variables
$\vec{A}_x$, and, by solving this system, we can obtain the inverse operators
in the form 
\begin{eqnarray}
U^\mathrm{inv}_x=\frac{1}{D^\mathrm{discr}_xD^\mathrm{cont}_x}(
A_x^\prime
+
B_x^\prime \theta_1
+
C_x^\prime \theta_2
+
D_x^\prime \theta_3
+
E_x^\prime \theta_{1}\theta_2
+
F_x^\prime \theta_{1}\theta_3
+
G_x^\prime \theta_{2}\theta_3
+
H_x^\prime \theta_{1}\theta_2\theta_3)\;,
\label{invOpDen}
\end{eqnarray}
where $D^\mathrm{discr}_x$ and $D^\mathrm{cont}_x$ are polynomials in the
discrete variables $a,b,c_1,c_2,c_3$ and the continuous variables $\vec{z}$,
respectively, and $\vec{A}_x^{\,\prime}$ are some rather cumbersome polynomials
in the variables of $F_C(\vec{z})$.
Specifically, we have
\begin{eqnarray}
D^\mathrm{discr}_a&=&\prod_{i\ne j=1 }^3(1+a-c_i)(2+a-c_i-c_j)
(3+a-c_1-c_2-c_3)\;,
\nonumber\\
D^\mathrm{discr}_b&=&D^\mathrm{discr}_a|_{a\to b}\;, 
\nonumber\\
D^\mathrm{discr}_{c_i}&=&\prod_{p\in\{a,b\}}\prod_{ j\ne i=1}^3
(1+p-c_i)(2+p-c_i-c_j)(3+p-c_1-c_2-c_3)\;.
\nonumber\\
\label{denomDiscr}
\end{eqnarray}
The denominator $D^\mathrm{cont}_a$ coincides with the surfaces of the
singularities of the PDE system for $F_C(\vec{z})$ with three variables,
\begin{eqnarray}
D^\mathrm{cont}_a&=&\left[-1+\sum_i^3( 3z_i-3 z_i^2+ z_i^3)-\sum_{i\ne j=1}^3 (z_i^2 z_j+z_i z_j )+10z_1z_2z_3\right](1-z_1-z_2-z_3)\;,
\nonumber\\
D^\mathrm{cont}_a&=&D^\mathrm{cont}_b=D^\mathrm{cont}_{c_i},\qquad  i=1,2,3\;.
\label{denomCont}
\end{eqnarray}

It is well known that, in the limit $z_i\to 0$, $F_C({\vec{z}})$ degenerates to
the Lauricella function of two variables,
$F_4(a,b,c_1,c_2,\vec{z})$.
So, by taking the limit $z_i\to 0$ in Eq.~(\ref{invOpDen}), we can obtain the
corresponding inverse operators \cite{hyperdire2} for $F_4$, with a reduced
number of independent functions,
\begin{equation}
U^\mathrm{inv}_x|_{z_i\to 0}
=\frac{1}{D^\mathrm{discr}_{i,x}D^\mathrm{cont}_{i,x}}(
A_{i,x}^\prime
+
\sum_{j\ne i=1}^3 B_{i,x}^\prime \theta_j
+
\sum_{j\ne i\ne k =1}^3 E_{i,j,x}^\prime \theta_j\theta_k)\:.
\label{invOpDenDeg}
\end{equation}
In Eq.~(\ref{invOpDenDeg}), the expressions for $D^\mathrm{discr}_{i,x}$ are
obtained from Eq.~(\ref{denomDiscr}) by putting all the factors involving the
variable $c_i$ to unity, and $D^\mathrm{cont}_{i,x}=1-\sum_{j\ne i=1}^3x_j$.

Using the explicit forms of the direct and inverse operators in
Eqs.~(\ref{dirAB}), (\ref{direct:FS}), and (\ref{invOpDen}) and eliminating
the higher powers of $\theta_i$ via the same procedure as in
Eq.~(\ref{invOpEqFin}), we can write the results of the differential
reduction according to Eq.~(\ref{reductionX}) in the following form:
\begin{eqnarray}
\lefteqn{F_C(a+n_1,b+n_2; c_1+m_1,c_2+m_2,c_3+m_3;\vec{z})}
\nonumber 
\\ 
&=&
\left[S_0(\vec{z})
+
\sum_{i} S_i(\vec{z}) \frac{\partial}{\partial z_i} 
+
\sum_{i\ne j} S_{ij}(\vec{z}) \frac{\partial^2}{\partial z_i \partial z_j}
+S_{123}(\vec{z}) \frac{\partial^3}{\partial z_1 \partial z_2\partial z_3}
\right]F_C(\vec{z})
\;,
\label{diffRed}
\end{eqnarray}
where $n_i$ and $m_i$ is a set of integers, and $S$, $S_j$, $S_{ij}$ are
polynomials in $z_i$ and the discrete variables of $F_C(\vec{z})$.

It is easy to see that, if one of the factors in the denominator
$D^\mathrm{discr}_x$ is equal to zero, we obtain from Eq.~(\ref{invOpEq}) some
new PDE identities,
\begin{eqnarray}
(
A_x^\prime
+
B_x^\prime \theta_1
+
C_x^\prime \theta_2
+
D_x^\prime \theta_3
+
E_x^\prime \theta_{1}\theta_2
+
F_x^\prime \theta_{1}\theta_3
+
G_x^\prime \theta_{2}\theta_3
+
H_x^\prime \theta_{1}\theta_2\theta_3)F_C(\vec{z})=0\;.
\label{invOpZero}
\end{eqnarray}
Eq.~(\ref{invOpZero}) means that the hypergeometric functions entering
Eq.~(\ref{Pfaff}) are expressible in terms of simpler hypergeometric
functions, e.g.\ Gauss hypergeometric functions, and corresponds to the
condition of reducibility of the monodromy group of $F_C(\vec{z})$.
As a consequence, the inverse operators in Eq.~(\ref{invOpDen}) and the
differential-reduction algorithm in Eq.~(\ref{diffRed}) can be expressed in a
simpler form involving just seven and six independent functions, respectively.

\boldmath
\section{FcFunction --- Mathematica-based program for the differential
reduction of the Lauricella function $F_C$}
\unboldmath

In this section, we present the Mathematica-based\footnote{%
This program package was tested using Mathematica~8.0 \cite{math}.}
program package {\bf FcFunction} for the differential reduction of the
Lauricella function $F_C(\vec{z})$ of three variables, which is freely
available from Ref.~\cite{bytev:hyper}.
%
It allows one to automatically perform the differential reduction in
accordance with Eq.~(\ref{diffRed}).
Its current version only handles non-exceptional parameter values.

The file {\tt readme.txt} provides a brief description of the installiation and
usage of the program package {\bf FcFunction}.
The main package file {\tt FcFunction.m} contains the general definitions of
the differential-reduction formulas.
All the cumbersome formulas needed for shifting the values of single
parameters are accommodated in additional files that are gzipped and end with
{\tt *.m.gz}.
The file {\tt example-FcFunction.m} includes the example calculations explained
in subsection~\ref{sec:ex}.

%

\subsection{Input format}

The program package {\bf FcFunction} may be loaded in the standard way:
$$
<< \mathrm{"FcFunction.m"}
$$
It includes the following basic routines for the Lauricella function
$F_C(\vec{z})$:
\begin{eqnarray}
{\bf FcIndexChange}[\mbox{changingVector}, \mbox{parameterVector}]
\label{input:f1}
\end{eqnarray}
and
\begin{equation}
{\bf FcSeries [\dots] },
\end{equation}
where ``${\rm parameterVector}$'' defines the list of parameters of that
function and ``${\rm changingVector}$'' defines the set of integers by
which the values of these parameters are to be shifted, i.e.\ the vector pairs
$(\vec{\gamma},\vec{\sigma})$ and $(\vec{k},\vec{l})$ in
Eq.~(\ref{reductionX}), respectively.
For example, the operator:
\begin{eqnarray}
{\bf FcIndexChange}[\{1,-1,0,0,2\},\{a,b,c_1,c_2,c_3,z_1,z_2,z_3\}]
\label{input:f4}
\end{eqnarray}
shifts the arguments of the function
$F_C(a,b;c_1,c_2,c_3;z_1,z_2,z_3)$ so as to generate
$F_C(a+1,b-1;c_1,c_2,c_3+2;z_1,z_2,z_3)$.

The function ${\bf FcSeries}[\ldots]$ is designed for the numerical evaluation
of $F_C(\vec{z})$ and its derivatives.
It returns the values of the Taylor series of $F_C(\vec{z})$ in
Eq.~(\ref{FC:series}) and its derivatives upon the commands:
\begin{eqnarray}
&&
{\bf FcSeries [ \mbox{vectorInit}, \mbox{numbSer}] } \;,
\nonumber\\\
&&
{\bf FcSeries}[\mbox{numberOfvariable}, \mbox{vectorInit}, \mbox{numbSer}] \;,
\label{cdDiffSeries}
\end{eqnarray}
respectively, where
``$\mbox{numberOfvariable}$'' is the list of the variables with respect to
which to differentiate,
``$\mbox{vectorInit}$'' is the set of parameters of $F_C(\vec{z})$, and
``$\mbox{numbSer}$'' is the number of terms to be retained in the Taylor
expansion.

\subsection{Output format}

The output structure of all the operators of the program package
{\bf FcFunction} in Eq.~(\ref{input:f1}) is as follows:
\begin{eqnarray}
\{\{Q_0,Q_1,Q_2,Q_3,Q_{12},Q_{13},Q_{23},Q_{123} \},\{ {\rm parameterVectorNew} \} \}\;,
\end{eqnarray}
where ``${\rm parameterVectorNew}$'' is the new set of parameters of
$F_C(\vec{z})$, i.e.\ $(\vec{\gamma}+\vec{k},\vec{\sigma}+\vec{l})$ in
Eq.~(\ref{reductionX}), and $Q_0, Q_1, Q_2,\ldots,Q_{123}$ are the rational
coefficient functions of the differential operators in Eq.~(\ref{invOp}), so
that
\begin{eqnarray}
F_C(\vec{\gamma};\vec{\sigma};\vec{z})&=&(Q_0+Q_1 \theta_1+Q_2\theta_2
+Q_3\theta_3+Q_{12}\theta_1\theta_2+Q_{13}\theta_1\theta_3
+Q_{23}\theta_2\theta_3
\nonumber\\
&&{}+Q_{123}\theta_1\theta_2\theta_3)
F_C(\vec{\gamma}+\vec{k};\vec{\sigma}+\vec{l};\vec{z})\;.
\end{eqnarray}

\subsection{Examples}
\label{sec:ex}

{\bf Example 1}:\footnote{%
All functions in the program package HYPERDIRE generate output without
additional simplification for maximum efficiency of the algorithm.
To get the output in a simpler form, we recommend to use the command
{\tt Simplify} in addition.}
Reduction of the Lauricella function $F_C(a,b;c_1,c_2,c_3;z_1,z_2,z_3)$.
\\
{\bf FcIndexChange[}\{$-1$,$0$,$1$,$0$,$0$\},\,\{$a$,$b$,$c_1$,$c_2$,$c_3$,$z_1$,$z_2$,$z_3$\}{\bf ]}
\\
\begin{eqnarray}
&&\left\{\left\{
 1-\frac{b z_1}{c_1 \left(z_1+z_2+z_3-1\right)} , \frac{\left(c_1-b\right) z_1+(a-1) \left(z_2+z_3-1\right)}{(a-1) c_1 \left(z_1+z_2+z_3-1\right)} , \frac{1-\frac{\left(a+b-c_2\right) z_1}{c_1
   \left(z_1+z_2+z_3-1\right)}}{a-1} ,
 \right.\right.
\nonumber\\
&&
\left.\left.
 \frac{1-\frac{\left(a+b-c_3\right) z_1}{c_1 \left(z_1+z_2+z_3-1\right)}}{a-1} ,
    \frac{-z_1+z_2+z_3-1}{(a-1) c_1 \left(z_1+z_2+z_3-1\right)} , \frac{-z_1+z_2+z_3-1}{(a-1)
   c_1 \left(z_1+z_2+z_3-1\right)} ,
 \right.\right.
\nonumber\\
&&-\left.\left.
\vphantom{\frac{1-\frac{\left(a+b-c_2\right) z_1}{c_1
   \left(z_1+z_2+z_3-1\right)}}{a-1}}
\frac{2 z_1}{(a-1) c_1 \left(z_1+z_2+z_3-1\right)} , 0  \right\},
\left\{ a-1 , b , c_1+1 , c_2 , c_3 , z_1 , z_2 , z_3
 \right\} \right\}
\end{eqnarray}
\\
In explicit form, this reads:
\begin{eqnarray}
\lefteqn{F_c(a,b;c_1,c_2,c_3;z_1,z_2,z_3)}
\nonumber\\
&=&\left[1
-\frac{b z_1}{c_1 \left(z_1+z_2+z_3-1\right)}
+ \frac{\left(c_1-b\right) z_1+(a-1) \left(z_2+z_3-1\right)}{(a-1) c_1 \left(z_1+z_2+z_3-1\right)}\theta_1
\right.
\nonumber\\
&&{}
+\frac{1-\frac{\left(a+b-c_2\right) z_1}{c_1\left(z_1+z_2+z_3-1\right)}}{a-1}\theta_2
+\frac{1-\frac{\left(a+b-c_3\right) z_1}{c_1 \left(z_1+z_2+z_3-1\right)}}{a-1}\theta_3
+\frac{-z_1+z_2+z_3-1}{(a-1) c_1 \left(z_1+z_2+z_3-1\right)}\theta_1\theta_2
\nonumber\\
&&{}+ \frac{-z_1+z_2+z_3-1}{(a-1)c_1 \left(z_1+z_2+z_3-1\right)}\theta_1\theta_3
-\left.\frac{2 z_1}{(a-1) c_1 \left(z_1+z_2+z_3-1\right)}\theta_2\theta_3
\right]
\nonumber\\
&&{}\times F_c (a-1,b;c_1+1,c_2,c_3;z_1,z_2,z_3)\;.
\label{example1}
\end{eqnarray}

The functions in Eq.~(\ref{cdDiffSeries}) allow us to expand the results as
formal Taylor series in the variables $z_i$ about zero and to analytically
check the results of the differential reduction in Eq.~(\ref{example1}).
For example, $F_C(a-1,b;c_1+1,c_2,c_3;z_1,z_2,z_3)$ and 
$\theta_1 \theta_2 F_C(a-1,b;c_1+1,c_2,c_3;z_1,z_2,z_3)$ may be Taylor expanded
through order ten as:
\begin{eqnarray}
&&{\bf FcSeries}[{a-1,b,c_1+1,c_2,c_3;z_1,z_2,z_3},10]\;,
\nonumber\\
&&{\bf FcSeries}[\{1,1,0\},{a-1,b,c_1+1,c_2,c_3;z_1,z_2,z_3},10]\:,
\end{eqnarray}
respectively.
Eq.~(\ref{cdDiffSeries}) is also useful for numerical estimations of the
Lauricella function $F_C(\vec{z})$ and its derivatives near the point
$\vec{z}=\vec{0}$.
However, the user has to control the convergence of the Taylor series and the
accuracy of the numerical evaluation.
Specifically, he has to ensure that the condition
$\sqrt{z_1}+\sqrt{z_2}+\sqrt{z_3}<1$ is satisfied.
Here are two examples:

\begin{eqnarray}
&&{\bf FcSeries[}\{1+\varepsilon,2+\varepsilon,4+3\varepsilon,6+7\varepsilon,3+3\varepsilon,0.1,0.2,0.15\},10{\bf ]}//.\varepsilon\rightarrow0.1
\nonumber\\
&&{\bf FcSeries[}\{1,0,0\}\{1+\varepsilon,2+\varepsilon,4+3\varepsilon,6+7\varepsilon,3+3\varepsilon,0.1,0.2,0.15\},10{\bf ]}//.\varepsilon\rightarrow0.1
\nonumber\\
&&1.34179
\nonumber\\
&&1.35774
\end{eqnarray}
\\

\noindent
{\bf Example 2}:
Reduction of the Lauricella function
$F_C(1+\varepsilon,1+2\varepsilon;3\varepsilon,4\varepsilon,5\varepsilon;
z_1,z_2,z_3)$.
\\
{\bf FcIndexChange[}\{$0$,$0$,$2$,$0$,$0$\},\,\{$1+\varepsilon$,$1+2\varepsilon$,$3\varepsilon$,$4\varepsilon$,$5\varepsilon$,z$_1$,z$_2$,z$_3$\}{\bf ]}
\\
\begin{eqnarray}
&&\!\!\!\!\!\left\{\left\{
1-\frac{(\epsilon +1) (2 \epsilon +1) z_1}{3 \epsilon  (3 \epsilon +1) \left(z_1+z_2+z_3-1\right)} ,
 \frac{3-\frac{z_1}{\epsilon  \left(z_1+z_2+z_3-1\right)}}{3 (3 \epsilon +1)} ,
  \frac{(\epsilon -3) z_1}{3\epsilon  (3 \epsilon +1) \left(z_1+z_2+z_3-1\right)} ,
\right.\right.
\nonumber\\
&&
\left.\left.
  \frac{(2 \epsilon -3) z_1}{3 \epsilon  (3 \epsilon +1) \left(z_1+z_2+z_3-1\right)} ,
  -\frac{2 z_1}{3 \left(3 \epsilon ^2+\epsilon \right)\left(z_1+z_2+z_3-1\right)},
\right.\right.
\nonumber\\
&&-\left.
\vphantom{\frac{3-\frac{z_1}{\epsilon  \left(z_1+z_2+z_3-1\right)}}{3 (3 \epsilon +1)}}
  \frac{2 z_1}{3 \left(3 \epsilon ^2+\epsilon \right) \left(z_1+z_2+z_3-1\right)} ,
       -\frac{2 z_1}{3 \left(3 \epsilon ^2+\epsilon \right) \left(z_1+z_2+z_3-1\right)},
    0
 \right\},
\nonumber\\
&&
 \left.\left\{
 \epsilon +1 , 2 \epsilon +1 , 3 \epsilon +2 , 4 \epsilon  , 5 \epsilon  , z_1 , z_2 , z_3
\right\}
\vphantom{\frac{3-\frac{z_1}{\epsilon  \left(z_1+z_2+z_3-1\right)}}{3 (3 \epsilon +1)}}
\right\}
\end{eqnarray}
\normalsize
\\
This corresponds to the following mathematical formula:
\begin{eqnarray}
\lefteqn{F_C(1+\varepsilon,1+2\varepsilon;3\varepsilon,4\varepsilon,
5\varepsilon;z_1,z_2,z_3)}
\nonumber\\
&=&\left[
1-\frac{(\epsilon +1) (2 \epsilon +1) z_1}{3 \epsilon  (3 \epsilon +1) \left(z_1+z_2+z_3-1\right)}
+\frac{3-\frac{z_1}{\epsilon  \left(z_1+z_2+z_3-1\right)}}{3 (3 \epsilon +1)}\theta_1
\right.
\nonumber\\
&&{}+\frac{(\epsilon -3) z_1}{3\epsilon  (3 \epsilon +1) \left(z_1+z_2+z_3-1\right)}\theta_2
+\frac{(2 \epsilon -3) z_1}{3 \epsilon  (3 \epsilon +1) \left(z_1+z_2+z_3-1\right)}\theta_3
\nonumber\\
&&{}-\frac{2 z_1}{3 \left(3 \epsilon ^2+\epsilon \right)\left(z_1+z_2+z_3-1\right)}\theta_1\theta_2
-\frac{2 z_1}{3 \left(3 \epsilon ^2+\epsilon \right) \left(z_1+z_2+z_3-1\right)}\theta_1\theta_3
\nonumber\\
&&{}-\left.\frac{2 z_1}{3 \left(3 \epsilon ^2+\epsilon \right) \left(z_1+z_2+z_3-1\right)}\theta_2\theta_3
   \right]
F_c(1+\varepsilon,1+2\varepsilon;2+3\varepsilon,4\varepsilon,5\varepsilon;
z_1,z_2,z_3)\;.\quad
\end{eqnarray}

\section{Conclusions}
\label{conclusion}

The differential-reduction algorithm \cite{SST} allows one to relate Horn-type
hypergeometric functions with parameters whose values differ by integers.
In this paper, we presented a further extension of the Mathematica-based
\cite{math} program package HYPERDIRE
\cite{hyperdire1,hyperdire2,hyperdire3,hyperdire4} for the differential
reduction of generalized hypergeometric functions to sets of basis functions by
including the Lauricella function $F_C(\vec{z})$ \cite{Lauricella} of three
variables.
We intend to complete the treatment of the Lauricella functions of three
variables in the future.

\section*{Acknowledgements}

We are grateful to M.Yu.~Kalmykov for fruitful discussions, useful remarks, and
valuable contributions to this paper.
The work of V.V.B. was supported in part by the Heisenberg--Landau Program.
This work was supported in part by the German Federal Ministry for Education
and Research BMBF through Grant No.\ 05H15GUCC1 and by the German Research
Foundation DFG through the Collaborative Research Centre No.~676
{\it Particles, Strings and the Early Universe---The Structure of Matter and
Space-Time}.

\end{document}